\documentclass[fleqn,12pt,twoside]{article}
\usepackage{espcrc1}

\usepackage{graphicx}

\newcommand{\mean}[1]{\left\langle #1 \right\rangle} 

\newcommand{\AmS}{{\protect\the\textfont2
  A\kern-.1667em\lower.5ex\hbox{M}\kern-.125emS}}

\title{Analysis of directed flow from three-particle correlations}

\author{N. Borghini\address[ULB]{Service de Physique Th{\'e}orique, CP225,
    Universit{\'e} Libre de Bruxelles, B-1050 Brussels}
  \thanks{Supported by ``Actions de Recherche Concert{\'e}es'' of 
    ``Communaut{\'e} Fran{\c c}aise de Belgique'' and IISN-Belgium},
  P. M. Dinh\address[SPhT]{Service de Physique Th{\'e}orique, CEA-Saclay, 
    F-91191 Gif-sur-Yvette cedex}
  and J.-Y. Ollitrault\addressmark[SPhT]}
       
\begin{document}

\maketitle

\begin{abstract}
We present a new method for analysing directed flow, based on a 
three-particle azimuthal correlation. 
It is less biased by nonflow correlations than two-particle methods, and 
requires less statistics than four-particle methods. 
It is illustrated on NA49 data. 
\end{abstract}

\section{INTRODUCTION}

The measurement of the successive harmonics of azimuthal correlations in a 
heavy ion collision \cite{Voloshin:1994mz},
$v_n \equiv \mean{e^{in(\phi-\Phi_R)}},$ where $\Phi_R$ is the impact 
parameter direction, is of paramount importance, since it yields information 
on the medium created in the collision. 
In particular, a novel behaviour has been predicted for the first harmonic 
$v_1$, the so-called {\em directed flow}, at ultrarelativistic energies 
\cite{Snellings:1999bt}. 

However, the analysis of directed flow at these energies is highly nontrivial, 
because $v_1$ is very small. 
Thus, methods based on an analysis of two-particle azimuthal 
correlations~\cite{Danielewicz:hn} are likely 
to be biased by ``nonflow'' two-particle 
correlations: quantum (HBT) correlations between identical particles and
correlations due to global momentum conservation have been shown to 
be important at SPS~\cite{Dinh:1999mn}, while correlations due
to minijets may be large at RHIC \cite{Kovchegov:2002nf}.
On the other hand, methods relying on four-particle correlations 
\cite{Borghini:2001vi}, which are free from this bias, are plagued by a lack 
of statistics due to the smallness of $v_1$, although they give good results 
for the analysis of elliptic flow $v_2$ (see Sec.~\ref{s:elliptic}). 
To remedy these shortcomings, we proposed a new method of $v_1$ analysis 
\cite{Borghini:2002vp}, based on the measurement of a {\em mixed} 
three-particle correlation, which involves both $v_1$ and $v_2$: 
\begin{equation}
\label{principe}
\mean{e^{i(\phi_a+\phi_b-2\phi_c)}}\simeq (v_1)^2 v_2,
\end{equation}
where $\phi_a$, $\phi_b$, and $\phi_c$ denote the azimuths of three particles 
belonging to the same event, and the average runs over triplets of particles 
emitted in the collision, and over events. 
Once $v_2$ has been obtained from a separate analysis, this equation yields
$(v_1)^2 v_2$, thus $v_1$. 

Here, we apply this method to NA49 data on Pb-Pb collisions at 158 $A$GeV. 
Results obtained using the ``standard'' flow analysis are given in 
Ref.\ \cite{A.Wetzler}. 
In our method, the first step in the analysis is the measurement 
of a reference $v_2$. 
Then, an equation analogous to Eq.~\ref{principe} 
yields an estimate of the {\em integrated} $v_1$, i.e., 
its average value over some phase space region. 
Finally, restricting $\phi_1$ in Eq.~\ref{principe} 
to a small $(p_T,y)$ bin allows one to obtain more detailed, 
{\em differential} measurements of $v_1$ as a function of 
transverse momentum or rapidity.

\section{ELLIPTIC FLOW FROM 2, 4, 6, 8-PARTICLE CORRELATIONS}
\label{s:elliptic}

As stated in the introduction, our method of analysis of directed flow $v_1$ 
requires the preliminary knowledge of an estimate of 
the elliptic flow $v_2$, integrated over some phase space region. 
Of course, this estimate must be obtained by analysing 
the same sample of events from which one wants to extract $v_1$. 

In practice, the average over phase space is a {\em weighted} average:
\begin{equation}
\mean{w_2 v_2}\equiv \mean{w_2\, e^{2i(\phi-\Phi_R)}},
\end{equation}
where $w_2$ is the chosen weight.
In order to reduce statistical fluctuations, $w_2$ must be larger 
for particles with stronger elliptic flow. 

The value of $\mean{w_2 v_2}$ is obtained using the cumulant method 
described in Ref.\ \cite{Borghini:2001vi}: 
one can extract estimates of $\mean{w_2 v_2}$ from cumulants of 
multi- (2-, 4-, 6-\ldots) particle correlations. 
 While two-particle methods are equivalent to the standard
flow analysis, higher orders are essentially free from nonflow 
effects. 
They were first used in analysing data obtained by the STAR
Collaboration at RHIC~\cite{Tang:2001yq}.

In Fig.\ \ref{fig:v2(b)}, we present application of the method 
to NA49 data, and show the dimensionless quantity 
\begin{equation}
\label{dimensionless}
v_2\equiv \frac{\mean{w_2 v_2}}{\sqrt{\mean{(w_2)^2}}},
\end{equation}
as a function of centrality for charged particles, where we have 
used $w_2=p_T$.  
We display estimates using cumulants of two-, four-, six-, and even 
eight-particle correlations \cite{Borghini:2001vi}, 
as well as the corresponding quantity for charged pions obtained 
from the ``standard'' subevent method \cite{Danielewicz:hn}. 

It is quite remarkable that the four-, six-, and eight-particle
estimates 
all agree: this supports the idea that they are indeed free from 
nonflow effects, and correspond to a genuine collective motion 
in the direction of the impact parameter. 
Moreover, these multiparticle estimates show a slight discrepancy 
with the two-particle values, as expected if nonflow correlations 
are sizable \cite{Borghini:2001vi}. 
Please note that the statistical uncertainties on high order 
cumulants remain reasonably small, especially for midcentral 
collisions. 
In the following, our reference elliptic flow value will preferably 
be the estimate from the four-particle cumulant, which is a priori 
the most reliable since it is free from nonflow correlations and 
has a smaller statistical error than higher order estimates. 

\section{INTEGRATED DIRECTED FLOW FROM 2, 3, 4-PARTICLE CORRELATIONS}

The next step in the analysis is to determine the average value of 
directed flow over some phase space region. As in the case of 
elliptic flow, we perform a weighted average
\begin{equation}
\mean{w_1 v_1}\equiv \mean{w_1\, e^{i(\phi-\Phi_R)}},
\end{equation}
where stronger weight is given to particles with stronger directed 
flow. 
In the NA49 analysis, we used a rapidity dependent weight 
$w_1=y-y_{\rm CM}$, where $y_{\rm CM}$ is the centre-of-mass rapidity. 

This weighted average is obtained from the following three-particle 
correlation:
\begin{equation}
\label{c3&flow}
\mean{w_1(a)w_1(b)w_2(c)e^{i(\phi_a+\phi_b-2\phi_c)}}
=\mean{w_1v_1}^2\,\mean{w_2v_2},
\end{equation}
where $w_2$ is the same as in Sec.~\ref{s:elliptic}. 
Using the value of $\mean{w_2 v_2}$ obtained in Sec.~\ref{s:elliptic}, 
we thus derive $\mean{w_1 v_1}$, up to a global sign. 

In practice, the left-hand side of Eq.~\ref{c3&flow} is 
constructed using a generating function formalism detailed in 
Ref.~\cite{Borghini:2002vp}. This procedure is a very efficient 
way to sum over all possible triplets of particles, and also 
to remove automatically the effects of slight detector anisotropies. 

\begin{figure}[htb]
\begin{minipage}[t]{0.48\linewidth}
\vspace{-10mm}
\includegraphics*[width=\linewidth]{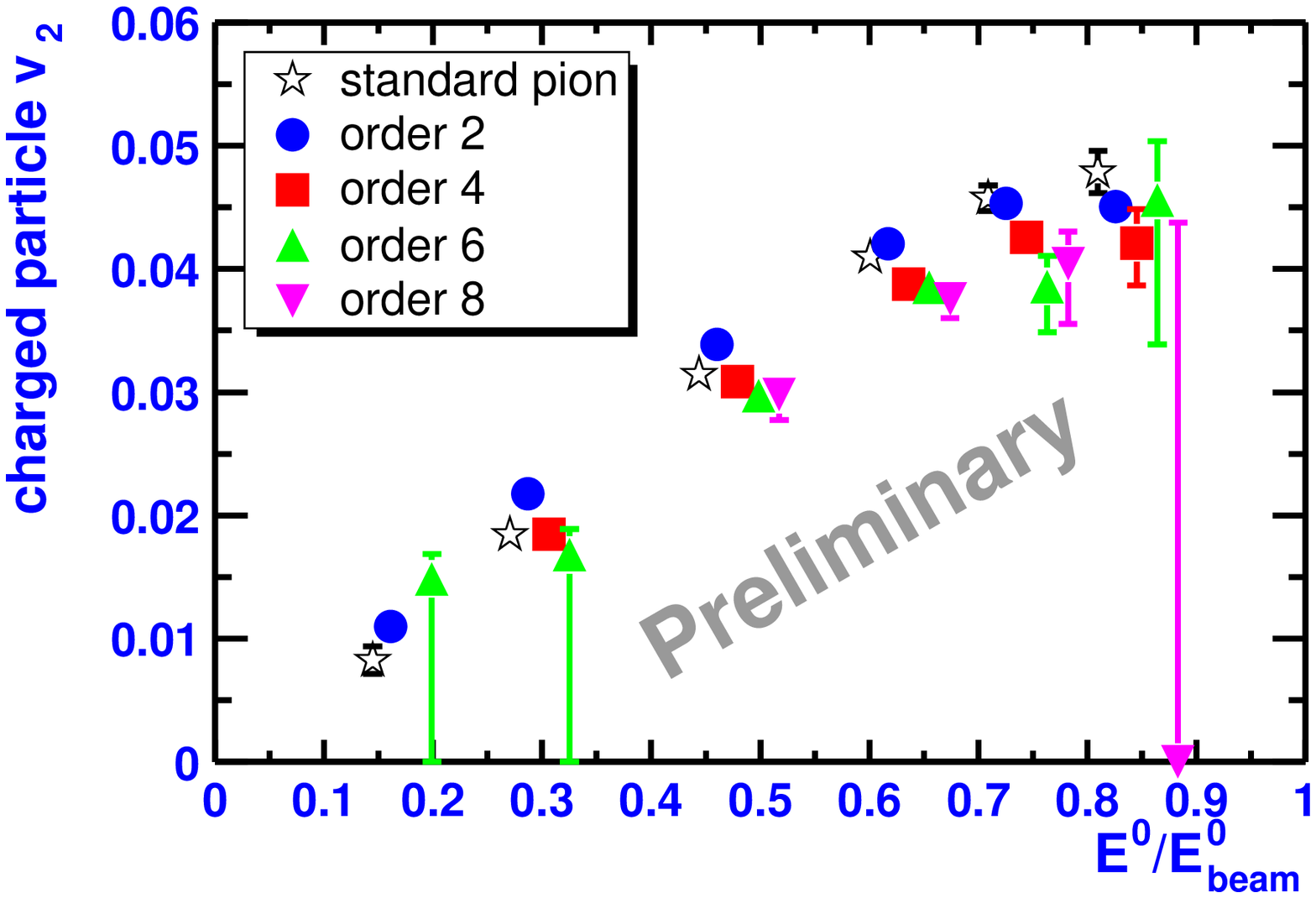}
\vspace{-10mm}
\caption{Weighted elliptic flow $v_2$ of charged particles as a function of 
centrality (central left, peripheral right) for Pb-Pb collisions at 158 
$A$GeV.}
\label{fig:v2(b)}
\end{minipage} \hspace{0.02\linewidth}
\begin{minipage}[t]{0.48\linewidth}
\vspace{-10mm}
\includegraphics*[width=\linewidth]{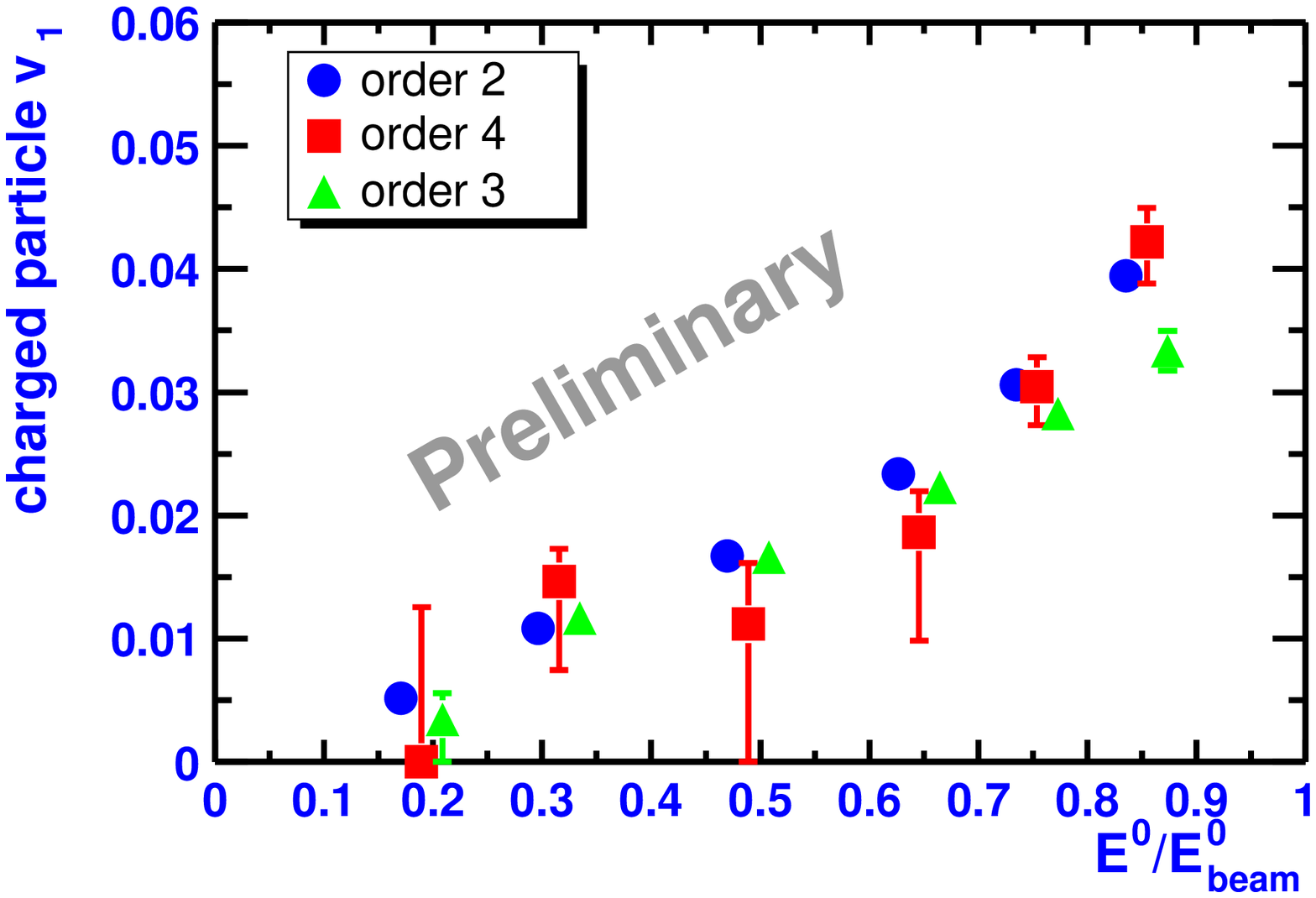}
\vspace{-10mm}
\caption{Weighted directed flow $v_1$ of charged particles as a function of 
centrality (central left, peripheral right).}
\label{fig:v1(b)}
\end{minipage}
\end{figure}

The method was applied to NA49 data. 
In Fig.\ \ref{fig:v1(b)}, we present as a function of centrality 
the dimensionless quantity 
$v_1\equiv \mean{w_1 v_1}/\sqrt{\mean{(w_1)^2}}$,  
analogous to Eq.~\ref{dimensionless}, 
together with two- and 
four-particle estimates obtained with the method of 
Ref.\ \cite{Borghini:2001vi}. 
It is worth noting that the statistical uncertainty on 
the three-particle estimate is 
barely larger than that on the two-particle value, while the systematic error 
due to nonflow correlations (not included in the plot) is a priori 
much smaller. 

\noindent
\vspace{-0.5cm}

\section{DIFFERENTIAL DIRECTED FLOW}

\begin{figure}[htb]
\begin{minipage}[t]{0.48\linewidth}
\vspace{-8mm}
\includegraphics*[width=\linewidth]{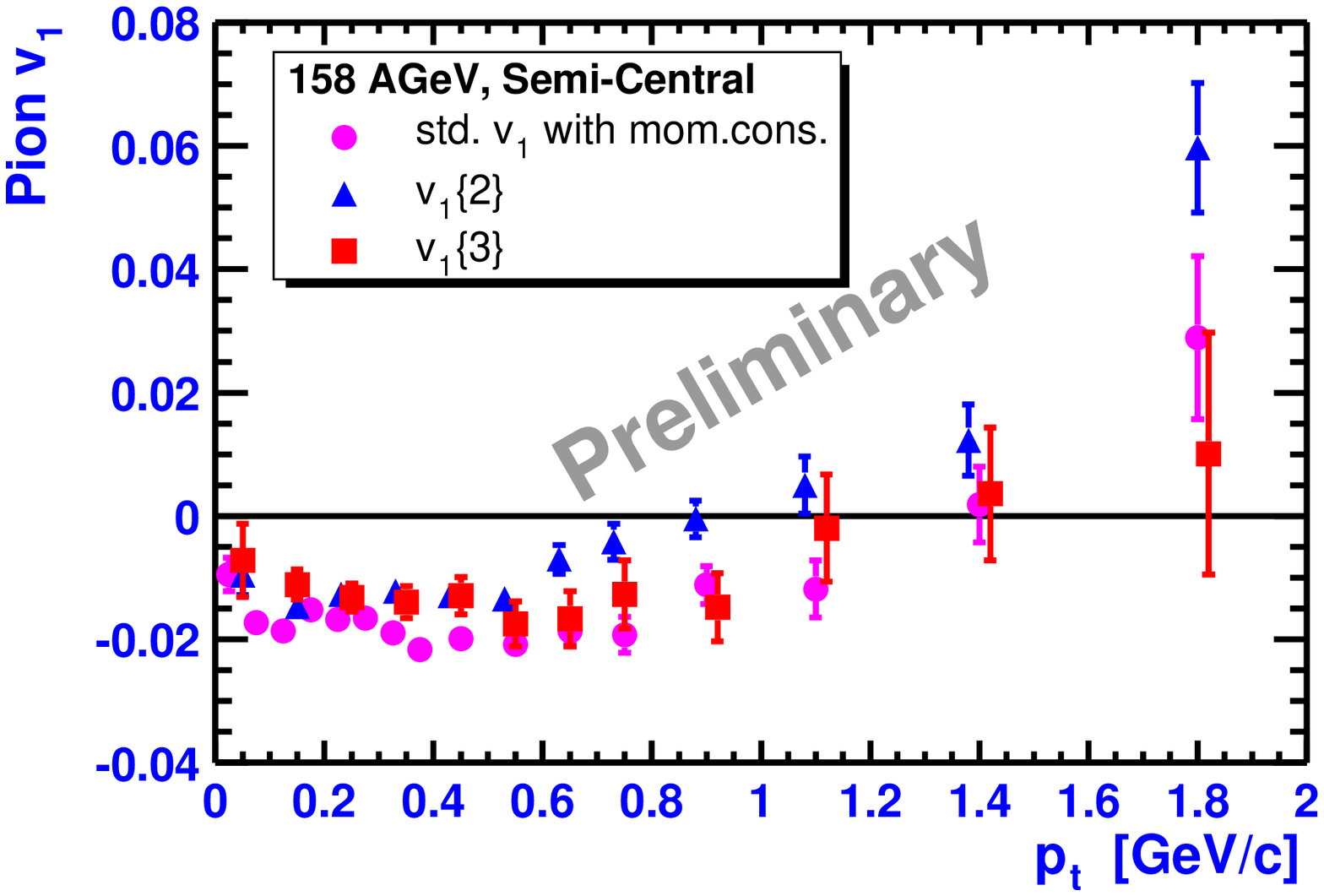}
\vspace{-10mm}
\caption{Directed flow $v_1$ of charged pions as a function of transverse 
  momentum for midcentral Pb-Pb collisions at 158 $A$GeV.}
\label{fig:v1(pT)}
\vspace{-8mm}
\end{minipage} \hspace{0.02\linewidth}
\begin{minipage}[t]{0.48\linewidth}
\vspace{-8mm}
\includegraphics*[width=\linewidth]{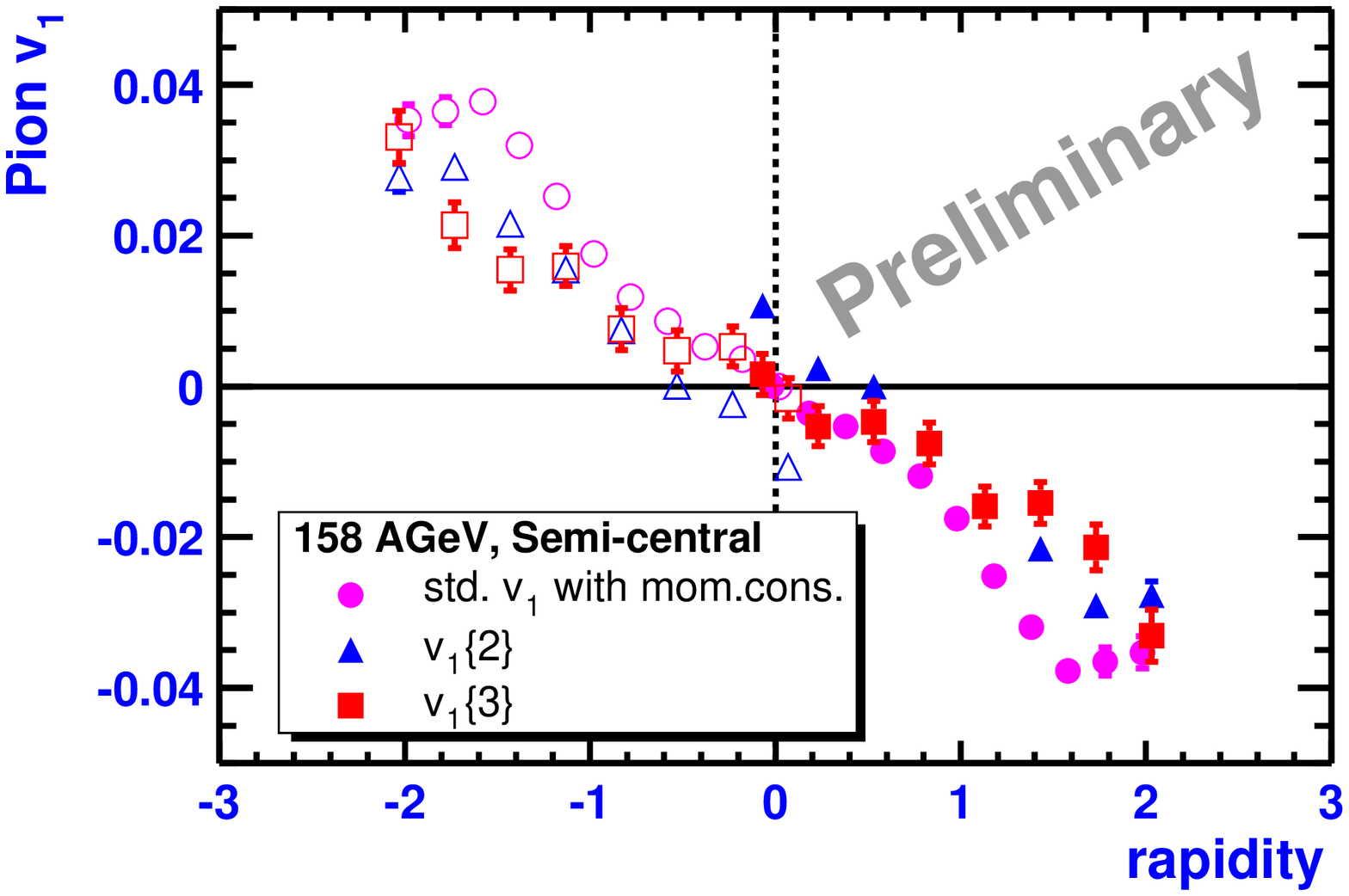}
\vspace{-10mm}
\caption{Directed flow $v_1$ of charged pions as a function of
$y-y_{\rm CM}$ for midcentral collisions. Open points are reflected 
with respect to midrapidity.}
\label{fig:v1(y)}
\vspace{-8mm}
\end{minipage}
\end{figure}

Finally, we want to obtain detailed measurements of $v_1$, as a function of 
$p_T$ or $y$ and for each particle type. 
To derive $v_1(p_T,y)$, we use a three-particle correlation 
analogous to Eq.~\ref{c3&flow}, where the particle labeled $a$ 
is restricted to a narrow $(p_T,y)$ bin:
\begin{equation}
\label{d3&flow}
\mean{w_1(b)w_2(c)e^{i(\phi_a+\phi_b-2\phi_c)}}
= \mean{w_1v_1}\mean{w_2v_2}v_1(p_{T\,a},y_a).
\end{equation}
With the previously derived $\mean{w_2v_2}$ and $\mean{w_1v_1}$, this 
equation yields $v_1(p_T,y)$. 
We illustrate this method on the differential flow of charged pions
in midcentral collisions, shown in 
Figs. \ref{fig:v1(pT)} and \ref{fig:v1(y)} together with two two-particle 
estimates, either uncorrected or corrected for the effect of momentum 
conservation~\cite{Borghini:2002mv}.

At high $p_T$, $v_1$ from three-particle correlations is consistent 
with the two-particle value corrected for momentum conservation, 
but significantly lower than the uncorrected one. This shows that 
correlations from momentum conservation, which are large, are 
automatically removed in our method. 
This is also reflected by the behaviour at midrapidity, 
where $v_1$ vanishes, as it should by symmetry, while the 
uncorrected two-particle estimate does not. 
The three-particle estimate also vanishes more smoothly at 
$p_T=0$ than two-particle estimates which may be biased by 
HBT correlations~\cite{Dinh:1999mn}.

\noindent

\section*{ACKNOWLEDGEMENTS}

We thank the NA49 Collaboration for permission to use their data.


\begin{thebibliography}{9}

\bibitem{Voloshin:1994mz}
S.~Voloshin and Y.~Zhang,
Z.\ Phys.\ C {\bf 70} (1996) 665. 

\bibitem{Snellings:1999bt}
R.~J.~M.~Snellings {\em et al.},
Phys.\ Rev.\ Lett.\  {\bf 84} (2000) 2803. 

\bibitem{Danielewicz:hn}
P.~Danielewicz and G.~Odyniec,
Phys.\ Lett.\ B {\bf 157} (1985) 146.

\bibitem{Dinh:1999mn}
P.~M.~Dinh, N.~Borghini and J.-Y.~Ollitrault,
Phys.\ Lett.\ B {\bf 477} (2000) 51; 

N.~Borghini, P.~M.~Dinh and J.-Y.~Ollitrault,
Phys.\ Rev.\ C {\bf 62} (2000) 034902. 


\bibitem{Kovchegov:2002nf}
Y.~V.~Kovchegov and K.~L.~Tuchin,
hep-ph/0203213.

\bibitem{Borghini:2001vi}
N.~Borghini, P.~M.~Dinh and J.-Y.~Ollitrault,
Phys.\ Rev.\ C {\bf 64} (2001) 054901; 
nucl-ex/0110016.

\bibitem{Borghini:2002vp}
N.~Borghini, P.~M.~Dinh and J.-Y.~Ollitrault,
Phys.\ Rev.\ C {\bf 66} (2002) 014905. 

\bibitem{A.Wetzler}
A.~Wetzler, these proceedings. 

\bibitem{Tang:2001yq}
A.~H.~Tang,
hep-ex/0108029;
C.~Adler {\it et al.}  [STAR Collaboration],
nucl-ex/0206001.

\bibitem{Borghini:2002mv}
N.~Borghini {\it et al.\/}, 
Phys.\ Rev.\ C {\bf 66} (2002) 014901.


\end{thebibliography}
\end{document}